\documentclass[12pt]{article}
\usepackage{graphicx}

\def \hf{\frac{1}{2}}

\def \bea{\begin{eqnarray}}
\def \beq{\begin{equation}}

\def \eea{\end{eqnarray}}
\def \eeq{\end{equation}}

\def \({\left(}
\def \){\right)}
\def \[{\left[}
\def \]{\right]}

\def \s{\sqrt{2}}

\def \ep{\epsilon}
\def \epstw {\epsilon^{(2)}}

\def \del {\delta}

\begin{document}
%
%
\rightline{arXiv:1311.1434}
\rightline{Phys. Lett. B 730, 221 (2014)}
\vskip 10mm
\centerline{\bf HIGH ORDER U-SPIN BREAKING:}
\medskip
\centerline{\bf A PRECISE AMPLITUDE RELATION IN $D^0$ DECAYS}
\bigskip
\centerline{Michael Gronau}
\medskip
\centerline{\it Physics Department, Technion -- Israel Institute of Technology}
\centerline{\it Haifa 3200, Israel}
\bigskip
\begin{quote}
U-spin breaking corrections up to third order are studied in $D^0$ decays to pairs 
involving a charged pion or kaon. The ratios $|A(D^0\to K^+\pi^-)|/|A(D^0 \to \pi^+K^-)|$ 
and $|A(D^0\to K^+K^-)|/|A(D^0\to \pi^+\pi^-)|$ determine values of $0.05$ and $0.30$
for real parts of two distinct first order U-spin breaking parameters of different origins.
We show that first and third order corrections vanish in the quantity
$\sqrt{|A(D^0\to K^+K^-)A(D^0\to \pi^+\pi^-)|}/\sqrt{|A(D^0\to K^+\pi^-)A(D^0 \to \pi^+K^-)|}=1$,
 while second order corrections cancel each other experimentally at a one percent 
level. We compare this ratio with the above two ratios and a third ratio involving these 
same four amplitudes, for which expansions up to and including second order are obtained. 
A nonlinear relation between these four ratios is shown to hold excluding third order 
U-spin breaking at a fraction of a percent. Isospin breaking in this relation and in the above 
equality  is suppressed by both isospin and U-spin breaking parameters. 
 \end{quote}
\bigskip

\section{Introduction\label{sec:introduction}}

U-spin symmetry, an SU(2) subgroup of flavor SU(3) under which the quark pair $(d,s)$
transforms like a doublet, has been shown to have powerful consequences in $D$ meson 
decays and in $D^0$-$\bar D^0$ mixing. Shortly after the 
discovery of charm in November 1974 a simple U-spin relation has been noted to hold 
among amplitudes for Cabibbo-favored (CF), singly 
Cabibbo-suppressed (SCS) and doubly Cabibbo-suppressed (DCS) $D^0$ 
decays~\cite{Kingsley:1975fe,Voloshin:1975yx},
\beq\label{U}  
A(\pi^+K^-):A(\pi^+\pi^-):A(K^+K^-):A(K^+\pi^-) = 
1:-\tan\theta_C:\tan\theta_C:-\tan^2\theta_C~,
\eeq
where $\theta_C$ is the Cabibbo angle. Early measurements observed that while 
$R_1\equiv |A(D^0\to K^+ \pi^-)|/$ $|A(D^0\to \pi^+K^-)|\tan^2\theta_C = 1$ holds within 
a reasonable approximation of order ten or twenty percent, the relation 
$R_2 \equiv |A(D^0\to K^+K^-)|/|A(D^0\to \pi^+\pi^-)| = 1$ is badly broken by about 
$80\%$. It has been recently suggested~\cite{Bhattacharya:2012ah,Brod:2012ud} that the large discrepancy of this ratio with respect to the U-spin symmetry value 
may be due to constructive interference between symmetry breaking in $\Delta U = 1$ 
``tree" and $\Delta U =0$ ``penguin" operators contributing to SCS decays, in contrast to the 
ratio of DCS and CF amplitudes which involves purely $\Delta U=1$ 
transitions~\cite{Savage:1991wu}. Understanding U-spin breaking in these decays may shed 
light on the relative strong phase 
$\delta$ between CF and DCS amplitudes, which vanishes in the U-spin 
symmetry limit~\cite{Kingsley:1975fe,Voloshin:1975yx,Wolfenstein:1995kv}. 
The phase $\delta$  plays a crucial role in determining $D^0$-$\bar D^0$ mixing 
parameters~\cite{Gronau:2000ru,Gronau:2001nr}, which formally vanish in the U-spin 
symmetry limit and also when including first order U-spin breaking~\cite{Gronau:2012kq}.

The purpose of this Letter is to examine the amplitude relations (\ref{U}) within the 
Cabibbo-Kobayashi-Maskawa (CKM) framework  when including first, second and third 
order U-spin breaking corrections. One of our motivations is searching for signals of new 
physics, which may be indicated by relations among amplitudes failing at some high 
order flavor symmetry breaking. Our study is also motivated by a very recent report of 
the LHCb collaboration ~\cite{Aaij:2013wda}, measuring the ratio of DCS and CF amplitudes 
$|A(D^0\to K^+\pi^-)|/|A(D^0 \to \pi^+K^-)|$ at an impressive high precision of less than a 
percent. Using this measurement we will update the status of second order corrections 
in a ratio $R_3$ of a sum of magnitudes of suitably normalized CF and DCS  amplitudes  and 
a sum of magnitudes of the two SCS amplitudes, in which first order U-spin breaking 
corrections have been suggesed to cancel~\cite{Brod:2012ud}. 
A new ratio $R_4$ involving products of amplitudes will be examined, in which
first and third order U-spin breaking corrections will be shown to vanish while second 
order corrections cancel experimentally at a one percent level. 
We will prove a nonlinear relation between $R_3 - R_4$ and $R_1$ and $R_2$, violated 
by a tiny third order correction - at most  a fraction of a percent. 
Isospin breaking corrections in this relation and in $R_3$ and $R_4$ will be shown 
to be suppressed by both isospin and U-spin breaking parameters.  Values of $R_1$ 
an $R_2$ will be used to calculate real parts of two distinct first order U-spin breaking 
parameters, ${\rm Re}\,\epsilon^{(1)} = 0.3$ and ${\rm Re}\,\epsilon^{(1)}_1 =0.05$. 
The imaginary part of the second parameter determines $\delta$. 

Studies assuming flavor SU(3) symmetry for $D$ and $D_s$ decays into all pairs of 
light pseudoscalar mesons have been presented in 
Refs.\,\cite{Bhattacharya:2008ss,Grossman:2012ry,Hiller:2012xm}. First order SU(3) 
breaking corrections were included in the latter two papers, identifying linear 
relations between amplitudes which hold in the presence of these corrections. 
Testing these linear relations involving at least three amplitudes requires knowledge of relative 
strong phases between amplitudes. One of these relations following from U-spin, involving 
the four amplitudes in (\ref{U}), was shown in Ref.\,\cite{Grossman:2012ry} to imply a corresponding relation among suitably normalized magnitudes of amplitudes as suggested 
in \cite{Brod:2012ud}.  Three other well-known amplitude relations  involving also a neutral 
pion or kaon follow from isospin symmetry~\cite{Kwong:1993ri}.

\section{U-spin symmetry limit\label{sec:Usymmetry}}

A formal proof of (\ref{U}) follows by considering U-spin properties of states and operators
denoted $|U, U_3\rangle$ and $(U, U_3)$, respectively. Initial $|D^0\rangle$ and final 
$|K^+K^- + \pi^+\pi^-\rangle/\s$ states are U-spin singlets $|0,0\rangle$, while  the three states, 
$-|\pi^+K^-\rangle, |K^+K^- - \pi^+\pi^-\rangle/\s, |K^+\pi^-\rangle$, are members of a triplet,
$|1,-1\rangle, |1,0\rangle, |1,+1\rangle$. The three pieces of the Hamiltonian operator responsible for CF, SCS and  DCS  decays behaving like $(\bar sd)$, $(\bar ss -\bar dd)$ and $(\bar d s)$ transform like a triplet:
\beq\label{H}
H_{\rm CF} = -\cos^2\theta_C(1,-1)\,,~~H_{\rm SCS} =\s \cos\theta_C\sin\theta_C(1,0)\,,~
H_{\rm\small DCS} = -\sin^2\theta_C(1,+1)\,.
\eeq
We use $V_{ud}=V_{cs} = \cos\theta_C,  V_{us}= - V_{cd} = \sin\theta_C$, neglecting in 
$H_{\rm SCS}$ tiny contributions proportional to $V^*_{cb}V_{ub}$, which 
may lead to small CP asymmetries at the level of $10^{-3}$ 
~\cite{Bhattacharya:2012ah,Brod:2012ud,Feldmann:2012js} but contribute negligibly 
to CP-averaged decay rates. The vanishing matrix element of a triplet operator  for the 
singlet final state $|K^+K^- + \pi^+\pi^-\rangle$ implies $A(\pi^+\pi^-)=-A(K^+K^-)$. Thus
the four amplitudes in (\ref{U}) are given in terms of a common U-spin triplet amplitude
$A \equiv \langle 1, U_3|(1, U_3)|0,0\rangle$,
\beq\label{U0}
A(\pi^+K^-) = \cos^2\theta_CA\,,~~~-\hskip-1mmA(\pi^+\pi^-) =A(K^+K^-)=\hf\sin 2\theta_CA\,,~
~A(K^+\pi^-) =\hskip-1mm-\sin^2\theta_CA\,,
\eeq
leading immediately to (\ref{U}).

\section{First, second and third order U-spin breaking\label{sec:highorder}}

We will introduce U-spin breaking corrections in (\ref{U}) assuming, as has been done 
in the 
past~\cite{Brod:2012ud,Savage:1991wu,Grossman:2012ry,Hiller:2012xm,
Feldmann:2012js,Hinchliffe:1995hz,Pirtskhalava:2011va,Gronau:2013mda}
that these corrections may be treated perturbatively.  Corrections of arbitrary order to 
decay amplitudes $\langle f|H_{\rm eff}|D^0\rangle$ are obtained by introducing in the 
Hamiltonian or in the final state powers of an $s$$-$$d$ spurion mass operator, 
$M_{\rm Ubrk} \propto (\bar s s) - (\bar d d) = \s(1,0)$. 
For SCS decays the effective Hamiltonian obtains at first order an additional $s+d$ penguin 
term $P_{s+d}$ due to an $s-d$ mass difference~\cite{Bhattacharya:2012ah}. That is, 
\beq\label{SCSfirst}
H_{\rm eff}M_{\rm Ubrk} =  H_{\rm SCS}M_{\rm Ubrk}  + P_{s+d}~,
\eeq
where the first term is a mixture of $(0,0)$ and $(2,0)$ while the second term behaves like a
pure U-spin singlet.
We will now show that corrections of given order have equal magnitudes in pairs of 
processes ($D^0\to \pi^+K^-$, $D^0 \to K^+\pi^-$) and ($D^0\to K^+K^-$, $D^0 \to \pi^+\pi^-$),
while their relative signs within these pairs are positive for even order and negative for odd order.
We will make a clear distinction between U-spin breaking parameters in CF or DCS 
decays and in SCS decays. 

Starting with first order corrections,
\beq\label{first}
\langle f|H_{\rm eff}|D^0\rangle^{(1)} = \langle f|H_{\rm eff}M_{\rm Ubrk}|D^0\rangle + 
\langle M_{\rm Ubrk}f|H_{\rm eff}|D^0\rangle~,
\eeq
we note that since the $D^0$ is a U-spin singlet 
only the triplet operators  in the products  $H_{\rm eff}M_{\rm Ubrk} \propto (1,\pm1)(1,0)$ 
contribute to the triplet final states $|f\rangle = |K^\pm\pi^\mp\rangle$, and only the triplet 
states in $M_{\rm Ubrk}|K^\pm \pi^\mp\rangle \propto (1,0)|K^\pm \pi^\mp\rangle$ obtain  
contributions from the triplet Hamiltonian operator. First order U-spin breaking corrections 
for $K^-\pi^+$ and $K^+\pi^-$ states, obtained by combining the two 
terms in (\ref{first}),  are equal in magnitude and have opposite signs  when leaving out prefactors 
$\cos^2\theta_C$ and $-\sin^2\theta_C$.  This sign change follows from an identity for 
Clebsch-Gordan coefficients, 
\beq\label{CG}
(n,0;1,-1|1,-1)= (-1)^n (n,0;1,1|1,1) = (1,1;n,0|1,1) = (-1)^n(1,-1;n,0|1,-1),
\eeq
applied to $n=1$. Denoting the correction parameter 
by $\epsilon_1^{(1)}$, where the {\em superscript represents the order in perturbation} and the 
{\em subscript marks the triplet nature of the transition operator}, one has
\beq\label{eps11}
\langle \pi^+K^-|H_{\rm eff}|D^0\rangle^{(1)} = -\cos^2\theta_CA\epsilon^{(1)}_1\,,~~~
\langle K^+\pi^-|H_{\rm eff}|D^0\rangle^{(1)} = -\sin^2\theta_CA\epsilon^{(1)}_1\,.
\eeq

First order U-spin breaking corrections for the triplet state $|K^+K^- - \pi^+\pi^-\rangle$ vanish
because of a vanishing Clebsch-Gordan coefficient, $(1,0;1,0|1,0)$ $=0$.
This implies that corrections for $K^+K^-$ and $\pi^+\pi^-$ have opposite signs
when leaving out prefactors $\cos\theta_C\sin\theta_C$ and $-\cos\theta_C\sin\theta_C$,
respectively. 
For the singlet state $|K^+K^- + \pi^+\pi^-\rangle$ one obtains two contributions
for the first term in (\ref{first}) originating in the two terms of $H_{\rm eff}M_{\rm Ubrk}$ in 
(\ref{SCSfirst}). The two terms correspond to a current-current (``tree") operator and an $s+d$
penguin operator occurring in the U-spin breaking phase.  (The second term in (\ref{first}) obtains 
only a tree contribution.) 
This distinguishes first order U-spin breaking in decays to $K^+K^-$ and 
$\pi^+\pi^-$ from that occurring in decays to $K^\pm\pi^\mp$ which,  as mentioned, involves only a triplet tree operator. Denoting by $\epsilon^{(1)}$ the correction parameter in the former decays, one has 
\beq\label{eps1}
\langle K^+K^-|H_{\rm eff}|D^0\rangle^{(1)} =
\langle \pi^+\pi^-|H_{\rm eff}|D^0\rangle^{(1)} = 
\cos\theta_C\sin\theta_CA\epsilon^{(1)}~.
\eeq

Second order U-spin breaking corrections are given by
\beq\label{second}
\langle f|H_{\rm eff}|D^0\rangle^{(2)} = \langle f|H_{\rm eff}M_{\rm Ubrk}^2|D^0\rangle + 
\langle M_{\rm Ubrk}^2f|H_{\rm eff}|D^0\rangle 
+ \langle M_{\rm Ubrk}f|H_{\rm eff}M_{\rm Ubrk}|D^0\rangle~,
\eeq
where $M^2_{\rm Ubrk} \propto (1,0)^2=-\sqrt{1/3}(0,0)$ $+ \sqrt{2/3}(2,0)$. 
For final states $|f\rangle = |K^\pm\pi^\mp\rangle$ we 
apply to the first two terms the Clebsch-Gordan identity (\ref{CG}) 
with $n=0$ and $n=2$ . Since in this case the identity involves no sign change, 
contributions of these two terms to $\pi^+K^-$ and $K^+\pi^-$ are equal in magnitudes 
and have equal signs when  leaving out the prefactors $\cos^2\theta_C$ and 
$-\sin^2\theta_C$. This is true also for the third term which involves squares of 
Clebsch-Gordan coefficients, as in this term the states $|M_{\rm Ubrk}f\rangle$ and 
$H_{\rm eff}M_{\rm Ubrk}|D^0\rangle$ must belong to the same U-spin representation.
Consequently,
\beq\label{eps12}
\langle \pi^+K^-|H_{\rm eff}|D^0\rangle^{(2)} = \cos^2\theta_CA\epsilon^{(2)}_1\,,~~~
\langle K^+\pi^-|H_{\rm eff}|D^0\rangle^{(2)} = -\sin^2\theta_CA\epsilon^{(2)}_1\,.
\eeq

Second order corrections for the singlet state involving $K^+K^-$ and  $\pi^+\pi^-$ 
vanish, $\langle K^+K^- + \pi^+\pi^-|H_{\rm eff}|D^0\rangle^{(2)} = 0$, because neither 
$U=0$ nor $U=2$ couples with $U = 1$ to $U=0$ (or vice versa) and
$(1,0;1,0|1,0) =0$. Thus,
\beq\label{eps2}
\langle K^+K^-|H_{\rm eff}|D^0\rangle^{(2)} =
-\langle \pi^+\pi^-|H_{\rm eff}|D^0\rangle^{(2)} = 
\cos\theta_C\sin\theta_CA\epsilon^{(2)}~.
\eeq
For the triplet state $|K^+K^- - \pi^+\pi^-\rangle$ the first and last terms in (\ref{second})
involve two contributions from  tree and $s+d$ penguin operators in 
$H_{\rm eff}M_{\rm Ubrk}$. As in the case of first order corrections, this distinguishes 
the second order parameter $\epsilon^{(2)}$ in decays to $K^+K^-$ and $\pi^+\pi^-$
from $\epsilon^{(2)}_1$ in $D^0 \to K^\pm\pi^\pm$ which is due to only triplet tree operators.

Third order corrections involve four terms,
\bea\label{third}
\langle f|H_{\rm eff}|D^0\rangle^{(3)} & = & \langle f|H_{\rm eff}M_{\rm Ubrk}^3|D^0\rangle + 
\langle M_{\rm Ubrk}^3f|H_{\rm eff}|D^0\rangle 
\nonumber\\
& + & \langle M_{\rm Ubrk}f|H_{\rm eff}M^2_{\rm Ubrk}|D^0\rangle
+ \langle M^2_{\rm Ubrk}f|H_{\rm eff}M_{\rm Ubrk}|D^0\rangle~,
\eea
where $M^3_{\rm Ubrk}$ is a mixture of $(1,0)$ and $(3,0)$. 
Applying an argument  similar to the one used for first order corrections and using the identity
(\ref{CG}), one may show that each of these four terms  
changes sign between $D^0\to \pi^+K^-$ and $D^0 \to K^+\pi^-$ and between 
$D^0 \to K^+K^-$ and $D^0 \to \pi^+\pi^-$
when leaving out $\theta_C$-dependent prefactors. Consequently, as in first order, the 
third order correction vanishes for the triplet state $|K^+K^- - \pi^+\pi^-\rangle$.
For the singlet state $|K^+K^- + \pi^+\pi^-\rangle$, all four terms in (\ref{third}) but the second
term involve contributions due to the two operators in (\ref{SCSfirst}), a current-current operator 
and an $s+d$ penguin operator.

Combining these properties of third order corrections with Eqs.\,(\ref{U0}), (\ref{eps11}),
(\ref{eps1}), (\ref{eps12}) and (\ref{eps2}) one obtains expressions 
for amplitudes including first, second and third order U-spin breaking corrections:
\bea\label{4amps}
A(D^0 \to \pi^+K^-) & = & \cos^2\theta_CA(1 - \epsilon^{(1)}_1 + \epsilon_1^{(2)} - 
\epsilon^{(3)}_1)~, \nonumber\\
A(D^0\to K^+\pi^-) & = & -\sin^2\theta_CA(1 + \epsilon^{(1)}_1 + \epsilon^{(2)}_1 + 
\epsilon^{(3)}_1)~, \nonumber\\
A(D^0 \to K^+K^-) & = & \cos\theta_C\sin\theta_CA(1 + \epsilon^{(1)} + \epsilon^{(2)} + 
\epsilon^{(3)})~, \nonumber\\
A(D^0 \to \pi^+\pi^-) & = & -\cos\theta_C\sin\theta_CA(1 - \epsilon^{(1)} + \epsilon^{(2)} - 
\epsilon^{(3)})~.
\eea
In each one of the two pairs of processes first and third order corrections 
occur with equal signs and may be combined into a single parameter, while the 
zeroth order term and the second order correction may be combined in
the first pair, changing the common factor by a second order correction.
This would provide expressions for the four complex amplitudes in terms of four complex
parameters,  which could be used for investigating U-spin breaking up to second order. 
We keep separately all six U-spin breaking parameters in (\ref{4amps}) in order to study 
up to third order one particular ratio involving these four amplitudes in which third order 
corrections vanish.  

We stress again that the two distinct U-spin breaking sets of parameters $\epsilon_1^{(n)}$ 
and $\epsilon^{(n)}$ ($n=1,2,3$) have different origins. While $\epsilon_1^{(n)}$ occur in CF 
and DCS decays, which are due to  pure $\Delta U=1$ tree  operators in $H_{\rm eff}$,  
$\epsilon^{(n)}$ in SCS decays combine  U-spin breaking in $\Delta U=1$ tree amplitudes 
with U-spin breaking in $\Delta U=0$ penguin operators with intermediate $s$ and $d$ 
quarks.  Consequently one naively expects $|\epsilon_1^{(1)}| \sim 0.2$ while $|\epsilon^{(1)}|$ 
may be considerably larger if these two U-spin breaking effects add up 
constructively~\cite{Bhattacharya:2012ah}. 
Higher order U-spin breaking parameters are expected to obey 
\beq
|\epsilon_1^{(n)}| \sim |\epsilon_1^{(1)}|^n~,~~~~~|\epsilon^{(n)}| \sim |\epsilon^{(1)}|^n~,
~~n=2,3~.
\eeq

An alternative notation for 
the U-spin breaking pattern (\ref{4amps}) could be in terms of two parameters
$\epsilon_1 \equiv \epsilon_1^{(1)}, \epsilon_2\equiv \epsilon^{(1)}$ and four coefficients
$a_i^n (i=1,2)$ of order one:
\bea\label{notation}
A(D^0 \to \pi^+K^-) & = & \cos^2\theta_CA[1 - \epsilon_1 + a^2_1(\epsilon_1)^2
 - a^3_1(\epsilon_1)^3]~, \nonumber\\
A(D^0\to K^+\pi^-) & = & -\sin^2\theta_CA[1 + \epsilon_1 + a^2_1(\epsilon_1)^2
 + a^3_1(\epsilon_1)^3]~, \nonumber\\
A(D^0 \to K^+K^-) & = & \cos\theta_C\sin\theta_CA[1 + \epsilon_2 + a^2_2(\epsilon_2)^2
 + a^3_2(\epsilon_2)^3]~, \nonumber\\
A(D^0 \to \pi^+\pi^-) & = & -\cos\theta_C\sin\theta_CA[1 - \epsilon_2 + a^2_2(\epsilon_2)^2
 - a^3_2(\epsilon_2)^3]~.
\eea
We will use the shorter notation (\ref{4amps}).

\section{Four ratios of amplitudes\label{sec:ratios}}

Magnitudes of amplitudes in (\ref{4amps}) may be expanded up to second order using 
\beq\label{expand}
|1 \pm \epsilon^{(1)} + \epsilon^{(2)}| = 1 \pm {\rm Re}\,\epsilon^{(1)} + 
\hf({\rm Im}\,\epsilon^{(1)})^2 + {\rm Re}\,\epsilon^{(2)} + {\cal O}[(\epsilon^{(1)})^3]~.
\eeq
Two often discussed ratios of amplitudes are:
\beq\label{R1}
R_1  \equiv \frac{|A(D^0 \to K^+\pi^-)|/|A(D^0 \to \pi^+K^-)|}{\tan^2\theta_C} 
\eeq
and 
\beq\label{R2}
R_2 \equiv \frac{|A(D^0 \to K^+K^-)|}{|A(D^0\to \pi^+\pi^-)|}~. 
\eeq
Using (\ref{4amps}) and (\ref{expand}) one obtains 
\bea\label{R12}
R_1 & = & 1 + 2[{\rm Re}\,\epsilon^{(1)}_1 + ({\rm Re}\,\epsilon^{(1)}_1)^2] + 
{\cal O}[(\epsilon^{(1)}_1)^3]~,
\nonumber\\
R_2 & = & 1 + 2[{\rm Re}\,\epsilon^{(1)} + ({\rm Re}\,\epsilon^{(1)})^2] +
 {\cal O}[(\epsilon^{(1)})^3]~.
\eea

These two ratios involve first order corrections given by $2{\rm Re}\,\epsilon_1^{(1)}$ 
and $2{\rm Re}\,\epsilon^{(1)}$. Interestingly {\em second order corrections in these ratios are 
given by squares of these same real parts} with no dependence on the second order 
parameters  $\epsilon_1^{(2)}$ and $\epsilon^{(2)}$. Thus measurements of $R_1$ and $R_2$ 
provide ways for calculating ${\rm Re}\,\epsilon^{(1)}_1$ and ${\rm Re}\,\epsilon^{(1)}$
up to third order corrections.
Eqs.\,(\ref{R12}) should include the U-spin symmetry limit, requiring solutions 
${\rm Re}\,\epsilon^{(1)}_1=0$ and ${\rm Re}\,\epsilon^{(1)}=0$ (rather than 
${\rm Re}\,\epsilon^{(1)}_1=-1$ and ${\rm Re}\,\epsilon^{(1)}=-1$) for $R_1=1$ 
and $R_2=1$, respectively.  This implies
\bea\label{Re}
{\rm Re}\,\epsilon_1^{(1)} & = & \hf\left(\sqrt{2R_1 - 1} -1\right) +  {\cal O}[(\epsilon^{(1)}_1)^3]~,
\nonumber\\
{\rm Re}\,\epsilon^{(1)} & = & \hf\left(\sqrt{2R_2 - 1} -1\right) + {\cal O}[(\epsilon^{(1)})^3]~.
\eea

A third ratio $R_3$ involving sums of amplitudes has been pointed out in 
Refs.\,\cite{Brod:2012ud,Grossman:2012ry} to differ from one by second order 
U-spin breaking corrections. Indeed, we find
\bea\label{R3}
R_3 & \equiv & \frac{|A(D^0 \to K^+K^-)| + |A(D^0\to \pi^+\pi^-)|}
{|A(D^0 \to \pi^+K^-)|\tan\theta_C + |A(D^0\to K^+\pi^-)|\tan^{-1}\theta_C} 
\nonumber\\
& = & 1 + \hf[({\rm Im}\,\epsilon^{(1)})^2 - ({\rm Im}\,\epsilon_1^{(1)})^2] + 
{\rm Re}\,(\epsilon^{(2)} - \epsilon^{(2)}_1) + {\cal O}[(\epsilon^{(1)})^3, (\epsilon_1^{(1)})^3]~.
\eea

We now propose to consider another ratio involving products of amplitudes,
\bea\label{R4}
R_4 & \equiv & \sqrt{\frac{|A(D^0 \to K^+K^-)||A(D^0\to \pi^+\pi^-)|}
{|A(D^0\to \pi^+K^-)||A(D^0\to K^+\pi^-)|}} 
\nonumber\\
& = & 1 - \hf{\rm Re}\,[(\epsilon^{(1)})^2 - (\epsilon_1^{(1)})^2] + 
{\rm Re}\,(\epsilon^{(2)} - \epsilon^{(2)}_1) + {\cal O}[(\epsilon^{(1)})^4, (\epsilon_1^{(1)})^4]
\nonumber \\
& = & 1 - \hf[({\rm Re}\,\epsilon^{(1)})^2 - ({\rm Re}\,\epsilon_1^{(1)})^2]
+ \hf[({\rm Im}\,\epsilon^{(1)})^2 - ({\rm Im}\,\epsilon_1^{(1)})^2] + 
{\rm Re}\,(\epsilon^{(2)} - \epsilon^{(2)}_1) 
\nonumber \\
& & + ~{\cal O}[(\epsilon^{(1)})^4, (\epsilon_1^{(1)})^4]~.
\eea
{\em Third order U-spin breaking corrections vanish in $R_4$ whereas they contribute 
in $R_3$.} These two ratios differ by a second order quantity, 
\beq\label{R3-4}
R_3 - R_4 = \hf[({\rm Re}\,\epsilon^{(1)})^2 - ({\rm Re}\,\epsilon_1^{(1)})^2]
+ {\cal O}[(\epsilon^{(1)})^3, (\epsilon_1^{(1)})^3]~,
\eeq
This and  Eqs.\,(\ref{Re}) lead to a relation between the four ratios of amplitude 
which holds up to and including second order U-spin breaking corrections,
\beq\label{sumrule}
R_4 = R_3 - \frac{1}{8}\left[(\sqrt{2R_2 - 1} - 1)^2  - (\sqrt{2R_1 - 1} -1)^2\right] 
+ {\cal O}[(\epsilon^{(1)})^3, (\epsilon_1^{(1)})^3]~.
\eeq   
This relation {\em which is not an identity} has an interesting consequence. 
$R_3$ involves a positive second order correction  
of about five percent. (A correction of $4.0\pm 1.6\%$, calculated 
in Ref.\,\cite{Brod:2012ud}  using earlier data, will be updated below to 
$5.6 \pm 0.8\%$ using more recent data.) The positive second order quantity 
$[(\sqrt{2R_2 - 1} - 1)^2  - (\sqrt{2R_1 - 1} -1)^2]/8$ is only around five percent 
in spite of the large U-spin breaking in $R_2$ because $(\sqrt{2R_2 - 1} - 1)^2/8$ 
involves a strong suppression of this correction while the contribution of the $R_1$ 
term is much smaller. Thus $R_4$ is very close to one; namely second order 
corrections in $R_4$ cancel each other.

\section{Numerical calculation of $R_i$ and ${\rm Re}\,\ep_1^{(1)}, 
{\rm Re}\,\ep^{(1)}$\label{sec:numerical}}
 
%
\begin{table}[h]
\caption{CP-averaged branching fractions~\cite{PDG} and amplitudes in units of 
$10^{-1}({\rm GeV}/c)^{-1/2}$ for $D^0$ decays to pairs involving a charged pion 
and kaon.\label{tab:BA}} 
\begin{center}
\begin{tabular}{c c c c} \hline \hline
Decay mode & Branching fraction (${\cal B}$)~\cite{PDG}& $p^*$ (GeV/c) & 
$|A|=\sqrt{{\cal B}/p^*}$ \\ \hline
 $D^0\to \pi^+K^-$  & $(3.88 \pm 0.05)\times 10^{-2}$ & $0.861$ & 
 $2.123 \pm 0.014$  \\ 
 $D^0 \to K^+\pi^- $ & $(3.88 \pm 0.05)\times 10^{-2}R_D$ & $0.861$ & 
 $0.1268 \pm  0.0014$ \\
 $D^0 \to K^+K^-$ & $(3.96 \pm 0.08)\times 10^{-3}$  & $0.791$ & 
 $0.708 \pm 0.007$ \\ 
 $D^0 \to \pi^+\pi^-$ &   $(1.402 \pm 0.026)\times 10^{-3}$ & $0.922$ & 
 $0.3899 \pm 0.0036$ \\ 
 \hline \hline
\end{tabular}
\end{center}
\end{table}

We now proceed to calculate the four ratios $R_i (i =1...4)$ using experimental data.
Table \ref{tab:BA} quotes CP-averaged branching fractions ${\cal B}$ for the four relevant 
decay processes, and magnitudes of amplitudes defined by $|A| \equiv \sqrt{{\cal B}/p^*}$. 
Since we are only concerned with ratios of amplitudes we disregard common phase space 
factors which cancel in these ratios. The current precision of the four  amplitudes is about one 
percent. Three of the four branching fractions are taken from Ref.\,\cite{PDG} while the fourth 
one is calculated using a very recent precise measurement of the ratio of DCS and CF  
branching fractions~\cite{Aaij:2013wda},
\beq
R_D \equiv \frac{{\cal B}(D^0 \to K^+\pi^-)}{{\cal B}(D^0 \to \pi^+K^-)} = 
(3.568 \pm 0.066)\times 10^{-3}~.
\eeq  

Using Cabibbo-Kobayashi-Maskawa  parameters\,\cite{PDG}, 
$\cos\theta_C = |V_{ud}| =0.97425 \pm 0.00022,$
$\sin\theta_C$ $= |V_{us}| = 0.2252 \pm 0.0009$ which imply $\tan\theta_C = 
0.2312 \pm 0.0009$, we calculate the following values for the four ratios:
\bea\label{Ri}
R_1 & = & 1.118 \pm 0.014~,
\nonumber\\
R_2 & = & 1.814 \pm 0.018~,
\nonumber\\
R_3 & = & 1.056 \pm 0.008~,
\nonumber\\
R_4 & = & 1.012 \pm 0.007~.
\eea
It is remarkable that second order U-spin breaking corrections in $R_4$ given
in Eq.\,(\ref{R4}) cancel each other at an accuracy of about one percent.

We note that the absolute branching fraction of $D^0\to \pi^+K^-$ and its error, for
which a new value was reported after completion of this work~\cite{Bonvicini:2013dda},
do not affect the central values and errors in $R_i$ because the other three branching 
fractions in Table \ref{tab:BA} have been measured by their ratios relative to this reference 
branching fraction~\cite{PDG}.  The latter three ratios determine $R_i$.
Thus the errors in $R_2, R_3$ ans $R_4$ calculated in 
(\ref{Ri}) are somewhat smaller than those which would have been obtained from 
errors in amplitudes given in Table \ref{tab:BA}.

Using Eqs.\,(\ref{Re}) we find
\bea\label{ReE11}
{\rm Re}\,\epsilon_1^{(1)} & = & 0.056 \pm 0.006 + {\cal O}[(\epsilon^{(1)}_1)^3]~,
\\
{\rm Re}\,\epsilon^{(1)} & = &  0.311 \pm 0.006 + {\cal O}[(\epsilon^{(1)})^3]~.
\eea
The  vastly different magnitudes of the real parts of the two U-spin breaking
parameters follow from the different origins of these parameters as explained
in Section \ref{sec:highorder}.  

Eq.\,(\ref{R3-4}) implies
\beq\label{R3-R4fromR12}
R_3 - R_4 = \hf[({\rm Re}\,\epsilon^{(1)})^2 - ({\rm Re}\,\epsilon_1^{(1)})^2]  
= 0.047 \pm 0.002~,
\eeq
where the the right-hand side is obtained from measured values of 
$R_1$ and $R_2$. This agrees extremely well with the central value 
of this difference calculated directly,
\beq\label{R3-R4}
R_3 - R_4 = 0.044~.
\eeq

\section{Isospin breaking}

We have observed a cancellation of second order U-spin breaking at a 
level of one percent in $R_4$,  and third order U-spin breaking at a fraction of a percent
in a nonlinear relation (\ref{sumrule}) between the four ratios $R_i$. At this high precision
one should also consider isospin breaking which is expected to be about one percent.

Isospin breaking is introduced in the Hamiltonian $H_{\rm eff}$ through a $d-u$ 
spurion mass operator,
$M_{\rm Ibrk} \propto (\bar d d - \bar u u)$, transforming like a combination of a U-spin 
singlet and triplet. Isospin breaking contributions of the U-spin singlet operator for the four
final states in (\ref{4amps})  may be absorbed into the U-spin symmetric amplitude $A$. 
Contributions of the triplet operator follow the signs of first order U-spin breaking corrections 
 in (\ref{4amps}), and are represented by two distinct parameters, $\delta_1$ - for U-spin 
triplet states $\pi^+K^-$ and $K^+\pi^-$, and $\delta_0$ - for $K^+K^-$ and $\pi^+\pi^-$, 
the two components of a U-spin singlet state. 

Instead of (\ref{expand}) we now expand:
\beq
|1 \pm \epsilon^{(1)} + \epsilon^{(2)} \pm \del_0 | = 1 \pm {\rm Re}\,\epsilon^{(1)} + 
\hf({\rm Im}\,\epsilon^{(1)})^2 + {\rm Re}\,\epstw \pm {\rm Re}\,\del_0 + 
{\rm Im}\,\epsilon^{(1)}{\rm Im}\,\del_0 + {\cal O}[(\epsilon^{(1)})^3]~.
\eeq
Consequently, $R_1$ and $R_2$ in (\ref{R12}) obtain additional isospin breaking terms,
$ 2{\rm Re}\,\del_1$ and $ 2{\rm Re}\,\del_0$, respectively, while $R_3$ and $R_4$ receive
an identical term, ${\rm Im}\,\epsilon^{(1)}{\rm Im}\,\del_0 - 
{\rm Im}\,\epsilon_1^{(1)}{\rm Im}\,\del_1$, involving both isospin and U-spin breaking.
Although this term cancels on the left-hand side of  (\ref{R3-4}), the right-hand side  
now obtains isospin breaking corrections of order $\epsilon^{(1)}\del_0$ and 
$\epsilon_1^{(1)}\delta_1$. Thus the nonlinear relation (\ref{sumrule}) involves new terms 
of this order which are suppressed by both isospin and U-spin breaking parameters.
This correction, expected to be about a fraction of a percent, is consistent with the tiny 
difference between the values calculated in (\ref{R3-R4fromR12}) and (\ref{R3-R4}).

\section{Conclusion\label{sec:conclusion}}

We have calculated first order U-spin breaking parameters ${\rm Re}\,\epsilon^{(1)}$ 
and ${\rm Re}\,\epsilon^{(1)}_1$ around 0.30 and 0.05 from $R_2$ and $R_1$ 
and small second order corrections in $R_3$ and $R_4$, at levels of five and 
one percent. The excellent agreement between (\ref{R3-R4fromR12}) and 
(\ref{R3-R4}) confirms the nonlinear relation (\ref{sumrule}), implying that third 
order U-spin breaking corrections in this relation are very tiny - at most a fraction 
of a percent. {\em These numbers and their hierarchy provide first evidence ever
justifying high order (i. e. up to and including third order) perturbative studies of U-spin 
breaking (or flavor SU(3) breaking) in $D$ meson decay amplitudes.}

Isospin breaking corrections in $R_3$, $R_4$ and in (\ref{sumrule}) have been 
shown to be suppressed by both isospin and U-spin breaking parameters and are 
expected to be at a level of a fraction of a percent, consistent with our numerical
calculations of $R_4$ and (\ref{sumrule}). {\em No flavor symmetry breaking effect down 
to this tiny level has been found which would indicate physics beyond the 
standard model.}

We wish to conclude with two remarks concerning open questions:
\begin{itemize}
\item The remarkable cancellation of second order U-spin breaking corrections 
in $R_4$ given in Eq.\,(\ref{R4}), and the vastly different magnitudes of 
${\rm Re}\,\epsilon^{(1)}_1$ and ${\rm Re}\,\epsilon^{(1)}$, seem to suggest a 
possible relation between first and second order U-spin breaking parameters,
${\rm Re}\,\epsilon^{(2)} = \hf{\rm Re}(\epsilon^{(1)})^2$. [${\rm Re}\,\epsilon_1^{(2)}$
and  $\hf{\rm Re}\,(\epsilon_1^{(1)})^2$ are expected to be very small in view of 
Eq.\,(\ref{ReE11}).] 
This could imply $a_2^2=1/2$ in the notation (\ref{notation}).
Although this may be a purely accidental cancellation,
one may seek an explanation for this relation.   
\item The result ${\rm Re}\,\epsilon_1^{(1)}=0.056 \pm 0.006$  determined by the ratio 
of amplitudes $|A(D^0\to K^+\pi^-)|/$ $|A(D^0\to \pi^+K^-)|$ is considerably 
smaller than typical U-spin breaking which is expected to be around $0.2-0.3$.
What does this imply for $\delta$, the relative strong phase between these CF and DCS 
decay amplitudes, a knowledge of which is required for determining 
$D^0$-$\bar D^0$ mixing parameters from time dependence in these decays? 

The phase $\delta$ vanishes in the U-spin symmetry limit and is given in the 
linear approximation
by $\tan\delta = -2{\rm Im}\epsilon_1^{(1)}$, which affects $R_3$ and $R_4$ 
quadratically but cannot be extracted from these observables.
We note however that in case the phase of $\epsilon_1^{(1)}$ is not very large 
or not very far from $180^\circ$, 
for instance making a modest assumption $|{\rm Arg}\,\epsilon_1^{(1)}| < 45^\circ$
or $|{\rm Arg}\,\ep^1_1 - 180^\circ| < 45^\circ$, 
the small value of ${\rm Re}\,\epsilon_1^{(1)}$ implies 
$|\delta| < 7^\circ$. This would determine $\delta$ at a much higher accuracy than 
achieved experimentally~\cite{Asner:2012xb,{Lu:2013qna}} using a method based 
on correlated production of $D^0$ and $\bar D^0$ in $e^+e^-$ 
collisions~\cite{Gronau:2001nr}. This point demonstrates the importance of 
understanding at least qualitatively or semi-quantitatively the phase of this U-spin 
breaking parameter. 
\end{itemize}

\section*{Acknowledgments}

I wish to thank Yuval Grossman and Jonathan Rosner for helpful comments and
 the referee for his suggestion to consider isospin breaking.

\end{document}